\documentclass[graphicx,superscriptaddress,longbibliography,prb,reprint]{revtex4-1}

\usepackage[version=3]{mhchem} 
\usepackage[T1]{fontenc}
\usepackage{siunitx}
\usepackage{graphicx}
\usepackage[utf8]{inputenc}

\usepackage[normalem]{ulem}
\useunder{\uline}{\ul}{}
\begin{document}
\author{Robin J. Dolleman}
\email{R.J.Dolleman@tudelft.nl}
\altaffiliation[Current affiliation: ]{2nd Institute of Physics, RWTH Aachen University, 52074 Aachen, Germany}
\affiliation{Kavli Institute of Nanoscience, Delft University of Technology, Lorentzweg 1, 2628 CJ, Delft, The Netherlands}
\title{Phonon Scattering at Kinks in Suspended Graphene}
\author{Yaroslav M. Blanter}
\affiliation{Kavli Institute of Nanoscience, Delft University of Technology, Lorentzweg 1, 2628 CJ, Delft, The Netherlands}
\author{Herre S. J. van der Zant}
\affiliation{Kavli Institute of Nanoscience, Delft University of Technology, Lorentzweg 1, 2628 CJ, Delft, The Netherlands}
\author{Peter G. Steeneken}
\affiliation{Kavli Institute of Nanoscience, Delft University of Technology, Lorentzweg 1, 2628 CJ, Delft, The Netherlands}
\affiliation{Department of Precision and Microsystems Engineering, Delft University of Technology, Mekelweg 2, 2628 CD, Delft, The Netherlands}
\author{Gerard J. Verbiest}
\email{G.J.Verbiest@tudelft.nl}
\affiliation{Department of Precision and Microsystems Engineering, Delft University of Technology, Mekelweg 2, 2628 CD, Delft, The Netherlands}

\begin{abstract}
Recent experiments have shown surprisingly large thermal time constants in suspended graphene ranging from 10 to 100 ns in drums with a diameter ranging from 2 to 7 microns. The large time constants and their scaling with diameter points towards a thermal resistance at the edge of the drum. However, an explanation of the microscopic origin of this resistance is lacking. Here, we show how phonon scattering at a kink in the graphene, e.g. formed by sidewall adhesion at the edge of the suspended membrane, can cause a large thermal time constant. This kink strongly limits the fraction of flexural phonons that cross the suspended graphene edge, which causes a thermal interface resistance at its boundary. Our model predicts thermal time constants that are of the same order of magnitude as experimental data, and shows a similar dependence on the circumference. Furthermore, the model predicts the relative in-plane and out-of-plane phonon contributions to graphene's thermal expansion force, in agreement with experiments. We thus show, that in contrast to conventional thermal (Kapitza) resistance which occurs between two different materials, in 2D materials another type of thermal interface resistance can be geometrically induced in a single material.
\end{abstract}
\maketitle

\section{Introduction}

The transport of phonons and heat in 2D materials like graphene \cite{geim2010rise} is essentially different from that in 3D materials, due to their large anisotropy between the in-plane and out-of-plane stiffness. This leads to extraordinary thermal properties, that have attracted much interest \cite{balandin2008superior,ghosh2008extremely,cai2010thermal,chen2010raman,nika2012two,faugeras2010thermal,xu2014length,lee2011thermal,dorgan2013high,chen2012thermalb,seol2010two,ghosh2010dimensional,pop2012thermal}. 
 Recently, we demonstrated a thermomechanical method \cite{PhysRevB.96.165421} to characterize the thermal time constant $\tau$ of suspended graphene membranes. We found that the values of $\tau$ are considerably larger than expected. Moreover, $\tau$ was found to scale with the diameter of the suspended drums, which could be explained by a model in which the transient heat transport is limited by a thermal boundary resistance. Several studies have shown that such a thermal interface resistance can emerge within the graphene due to grain boundaries \cite{azizi2017kapitza,cao2012kapitza}, carbon isotope doping \cite{pei2012carbon}, encasing with boron nitride \cite{xu2014interfacial}, a step in the substrate \cite{sevinccli2014phonon} or a change in the number of graphene layers \cite{rojo2018thermal}. However, none of these microscopic models predict the emergence of a sufficiently large thermal boundary resistance to account for the large thermal time constants observed in Ref. \onlinecite{PhysRevB.96.165421}.

\begin{figure}
\includegraphics{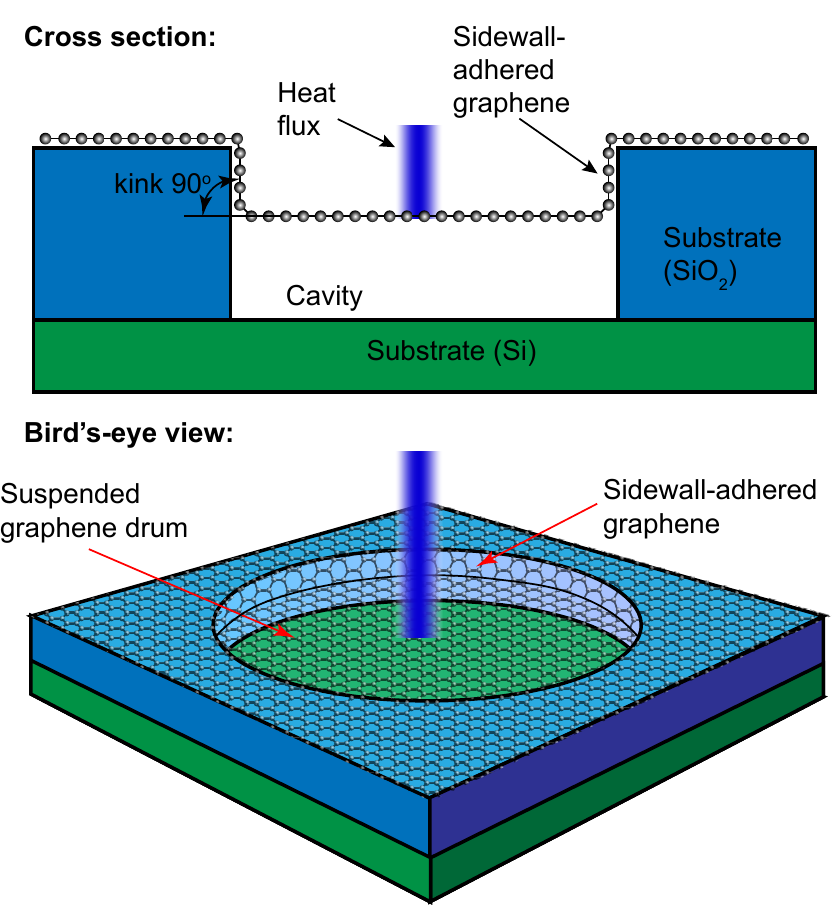}
\caption{Model system under consideration in this work. A graphene drum is suspended over a cavity and adheres to the sidewall, introducing a kink at the edge of suspended graphene.  The suspended graphene is heated by a laser and heat flow in the suspended graphene is studied. \label{fig:system}}
\end{figure}

Here, we theoretically analyze phonon transport in suspended graphene membranes and compare this to experimental works \cite{PhysRevB.96.165421,noncontinuum} on devices as depicted in Fig. \ref{fig:system}, to explain the large values of the thermal time constants. A laser heats up the center of the membrane, and the resulting heat is transported by lattice vibrations (phonons) to the substrate. It is well known that suspended 2D materials usually show a kink at their edge due to sidewall adhesion \cite{doi:10.1021/nl801457b,doi:10.1063/1.3270425,bunch2012adhesion}. 
For phonons to leave the suspended membrane, they have to be transmitted across the kink between the suspended and supported graphene. We show that this transmission is very small for flexural phonons, which is related to their low propagation speed compared to the in-plane phonons. Consequently, a thermal  interface resistance can arise in 2D materials from a kink within the material itself, even when the acoustic properties on both sides of the interface are equal. 
 The model predicts thermal time constants $\tau_{\mathrm{ZA}}$ in line with the experimental values found in Ref. \onlinecite{PhysRevB.96.165421}. 

The remainder of this article is structured as follows; section \ref{sec:mechmodel} constructs the mechanical model to calculate the transmission and reflection coefficients of a phonon incident on a kink.  In section \ref{sec:tauZA}, we use the mechanical modal as a boundary condition to construct a two-temperature model from which the thermal time constants and their relation to the thermal expansion forces can be calculated. In section \ref{sec:disc} we discuss how the model could be improved and make suggestions for future experiments. Finally, the conclusions are presented in section \ref{sec:concl}.

\section{Mechanical model for a kink}\label{sec:mechmodel}
To examine the effect of kinks in graphene on phonon transport, we develop a mechanical model that evaluates the phonon scattering at a kink with an angle $\beta$ and gives the phonon transmission and reflection probabilities. Figure~\ref{fig:kinkfig} shows that after an acoustic phonon reaches an interface, it will be converted in a combination of reflected and transmitted longitudinal (LA), transverse (TA) and flexural (ZA) acoustic phonons.
We find the transmission and reflection coefficients for each incident phonon mode by solving 6 coupled equations: 3 from the continuity of displacement and 3 from the continuity of stress. The derivation follows the method by Kolsky \cite{kolsky1963stress} closely, with additions to include the effects of the flexural phonons. To simplify the analysis, the second kink between the supported and sidewall-adhered graphene is not taken into account and all of the supported graphene and the substrate is assumed to be an ideal heat sink. In order to only observe the geometry induced effects of the kink, we set the elasticity parameters and tension equal in both domains, resulting in equal propagation velicities for each phonon mode on the suspended and supported graphene. 

\subsection{Snell's law}
\begin{figure}[t!]
\centering
\includegraphics{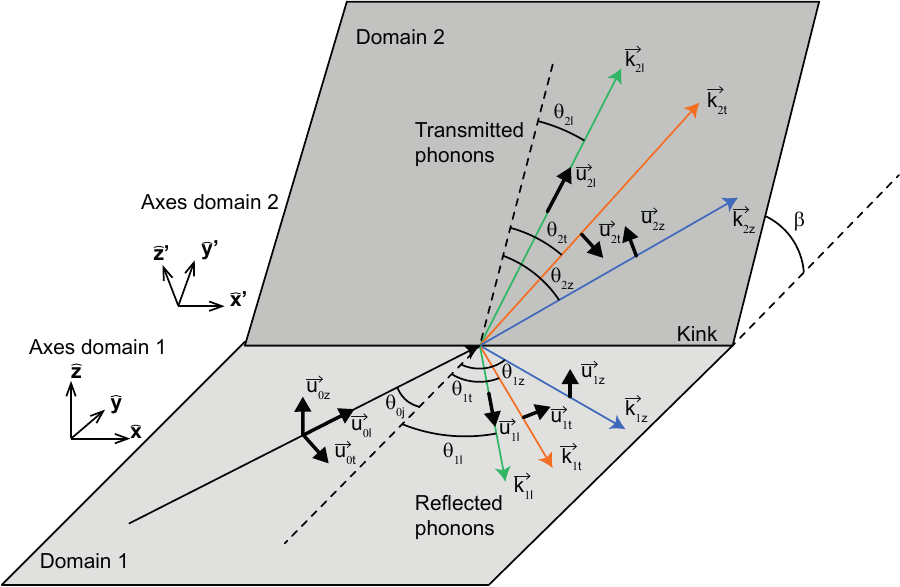}
\caption{Phonon scattering on a kink with angle $\beta$ in graphene. A phonon with amplitude $u_{0j}$ is incident on the interface with an angle $\theta_{0j}$, and the interface consists of a sharp kink in the graphene with angle $\beta$. The incident phonon can scatter into 6 possibilities, either transmission at LA, TA or ZA phonon or reflection as a LA, TA or ZA phonon. \label{fig:kinkfig}}
\end{figure}
The model calculates the transmission coefficients $w_{ij \rightarrow qr}$, which represent the fraction of phonons in mode $j$ on domain $i$ that reach the kink and end up into phonon mode $r$ on domain $q$. Here, we use $j,q = l, t, z$ for LA, TA and ZA phonons, respectively, and  $i,r = 1, 2$ for suspended or supported graphene, respectively. Also, the subscript $i=0$ is used to indicate an incident phonon from domain 1. 
 We consider the reflection and transmission of an incident phonon with amplitude $\vec{u}_{0j}$ and with an incident angle $\theta_{0j}$ (Fig. \ref{fig:kinkfig}), that is incident on an interface where the graphene has a kink with angle $\beta$. If the phonon propagation speed $c_{ij}$ is known, we can find the angles of reflection and refraction with respect to the normal using Snell's law:
\begin{equation}\label{eq:snellslaw}
\sin{\theta_{ij}} = \frac{c_{ij}}{c_{0j}} \sin{\theta_{0j}}.
\end{equation}
With the angles of refraction known, only the amplitudes $\vec{u}_{ij}$ of the reflected and refracted waves are unknown. To find these, we construct 6 coupled equations in the following subsections. 

\subsection{Continuity of deflection}
The mechanical motion $\vec{q}_{ij}$ around the static position of the membranes is described by a wave with amplitude $\vec{u}_{ij}$:
\begin{equation}\label{eq:wave}
\vec{q}_{ij} (x,y,t) = \vec{u}_{ij}\cos(\omega t + {k}_x x + {k}_y y),
\end{equation}
where ${k}_x$ is the component of the wavevector $\vec{k}$ in the $x$ direction of the local axis and ${k}_y$ in the $y$ direction. 
Positive directions of the displacements and wavevectors are defined as drawn in Fig. \ref{fig:kinkfig}.
The displacements in domain 2 are projected onto the coordinate system of domain 1, which gives 3 expressions for the continuity of displacement at the interface:
\begin{equation}\label{eq:contdefl}
\begin{aligned}
\sum\limits_{j}^{~}  \vec{q}_{1j}= \sum\limits_{j}^{~}  \vec{q}_{2j},\\
\end{aligned}
\end{equation}
By substituting Eq. \ref{eq:wave} in Eq. \ref{eq:contdefl}, and setting the origin $x = y = z =t=0$ to the location and time where the phonon hits the kink, one obtains expressions that only depend on the amplitudes $\vec{u}_{ij}$, the angles $\theta_{ij}$ and $\beta$. The full expressions are presented in the Supplemental Information S1 \cite{supplemental}. 

\subsection{Continuity of stress}
\begin{figure}[t!]
\includegraphics{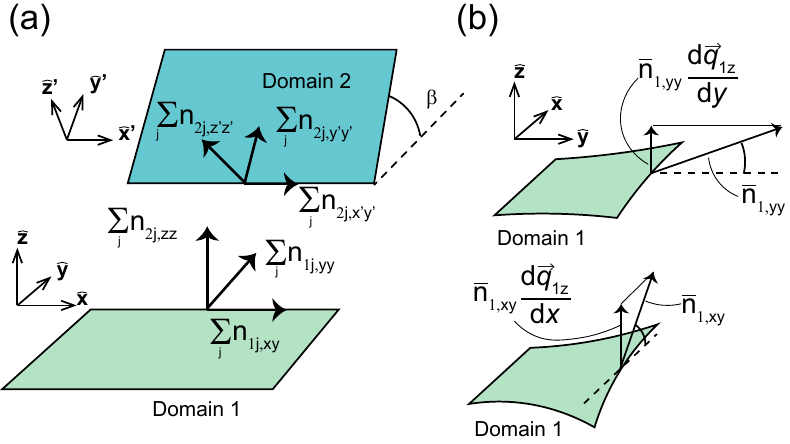}
\caption{Cross-section of the membrane showing the stresses at the interface. (a) The three stress components at each interface for both domains, which are added together in the axes of domain 1 to obtain the continuity of stress equations. (b) The out-of-plane displacement of the membrane $\vec{q}_{iz}$ results in a out-of-plane projection of the in-plane stress and shear components.\label{fig:stress}}
\end{figure}
The continuity of stress implies that the total tension is equal on both sides of the interface. Figure \ref{fig:stress} shows the relevant tension components at the interface, where ${n}_{ij,yy}$ and ${n}_{ij,zz}$ are the tension components in the $\hat{\textbf{y}}$, $\hat{\textbf{z}}$ directions, respectively, and ${n}_{ij,xy}$ is the shear stress component. Note, that ${n}_{ij,xx}$ does not play a role in the transmission of elastic waves because of rotational symmetry along the $\hat{\textbf{x}}$-direction. Furthermore, the components $n_{ij,xz} = n_{ij,yz} = 0$ due to the two-dimensional nature of the material.  Each remaining tension component of the tension tensor ${n}$ is then split into a static part $\overline{{n}}$ and a dynamic part $\delta {n}$ (for example: $n_{ij,yy}(t) = \overline{{n}}_{i,yy} + \delta {n}_{ij,yy}(t)$; the static component cannot be attributed to a specific phonon mode and therefore the subscript $j$ is omitted). To formulate the continuity of stress equations we only take the dynamic stress components into account, since the equilibrium is already satisfied for the static part of the stress. 

The dynamic stress components $\delta {n}_{ij,yy}$ and $\delta {n}_{ij,xy}$ are related to the deflection-induced dilatation and shear of the lattice by the relations \cite{kolsky1963stress}:
\begin{equation}\label{eq:nyyij}
\delta {n}_{ij,yy} = (\lambda + 2 \mu)\frac{\mathrm{d}\vec{q}_{ij}}{\mathrm{d}y}\hat{\textbf{y}}  + \lambda \frac{\mathrm{d} \vec{q}_{ij}}{\mathrm{d} x}\hat{\textbf{x}},
\end{equation}
\begin{equation}\label{eq:nxyij}
\delta {n}_{ij,xy} = \mu \frac{\mathrm{d} \vec{q}_{ij}}{\mathrm{d} y}\hat{\textbf{x}} + \mu \frac{\mathrm{d} \vec{q}_{ij}}{\mathrm{d} x}\hat{\textbf{y}},
\end{equation}
where $\lambda$ and $\mu$ are the Lame parameters; note, that these components are expressed in the local axes of each domain. The dynamic component $\delta {n}_{iz,zz}$ is a result of the flexural phonons, whose out-of-plane motion allows the static in-plane stress components $\overline{{n}}_{i,yy}$ and $\overline{{n}}_{i,xy}$ to be rotated into the $\hat{\textbf{z}}$-direction of the local axes, as shown in Fig. \ref{fig:stress}(b). The out-of-plane deflections $\vec{q}_{iz}$  are assumed to be small enough to not introduce significant dynamic tension modulations due to elastic deformation compared to the static pre-tension. This gives for the tension modulation component $\delta {n}_{iz,zz}$ in the local axis of each domain:
\begin{equation}\label{eq:nzziz}
\begin{aligned}
\delta {n}_{iz,zz} = \overline{{n}}_{i,yy} \frac{\mathrm{d} \vec{q}_{iz}}{\mathrm{d} y} \hat{\textbf{y}}+  \overline{{n}}_{i,xy}    \frac{\mathrm{d} \vec{q}_{iz}}{\mathrm{d}x} \hat{\textbf{x}}.
\end{aligned}
\end{equation}
By substituting Eq. \ref{eq:wave} into Eqs. \ref{eq:nyyij}--\ref{eq:nzziz}, the stress components shown in Fig. \ref{fig:stress} can be calculated and projected onto each of the axes of domain 1:
\begin{equation}\label{eq:contstress}
\begin{aligned}
\sum\limits_{\alpha \zeta}^{~} \sum\limits_{j}^{~}  \delta {n}_{ij,\alpha \zeta} \hat{\textbf{s}}_{\gamma}=
\sum\limits_{\alpha \zeta}^{~}  \sum\limits_{j}^{~} \delta {n}_{ij,\alpha \zeta} \hat{\textbf{s}}_{\gamma}, 
\end{aligned}
\end{equation}
where $\alpha \zeta \in \{xy, yy, zz,x'y',y'y',z'z'\}$, $\gamma \in \{x, y, z\}$ and $\hat{\textbf{s}}_{\gamma}$ is a unit vector pointing in one of the directions of domain 1.  
This results in three expressions that only depend on $\vec{u}_{ij}$, $\theta_{ij}$, $\beta$ and the pre-tension components $\bar{n}$, which are shown in the Supplemental Information S1 \cite{supplemental}. 

\subsection{Integrated Transmission Coefficients}
 The 6 equations we derived (Eqs. \ref{eq:contdefl} and \ref{eq:contstress}) can be solved simultaneously for each incident mode, by setting $|u_{0j}| = 1$ (see Supplemental information S1 for more details \cite{supplemental}). From the amplitudes of the transmitted and reflected waves, one can calculate the energy flux of each wave leaving the kink ($B_{ij} = \rho \omega^2 c_{ij} |u_{ij}|^2 \mathrm{Re}(\cos{\theta_{ij}})$, where $\rho$ is the density of graphene and $\omega$ the phonon frequency) and from that define the transmission coefficient as \cite{peterson1973kapitza}:
\begin{equation}\label{eq:transmissionprobmodel}
w_{0j\rightarrow qr} (\theta_{0j})= \frac{B_{qr}}{B_{0j}} = \frac{c_{qr} |u_{qr}|^2 \mathrm{Re}(\cos{\theta_{qr}})}{c_{0j} |u_{0j}|^2  \cos{\theta_{0j}}},
\end{equation}
where the incoming wave amplitude $|u_{0j}| = 1$. Note, that the density $\rho$ drops out of the equation because it is equal on both domains. In the model $w_{0j\rightarrow qr} (\theta_{0j})$ is integrated over all incoming angles $\theta_{0j}$ to obtain the total transmission or reflection coefficient of each scattering process $\bar{w}_{0j \rightarrow qr}$. $\bar{w}_{0j \rightarrow qr}$ can then be used to calculate the total heat flux crossing the boundary.   However, we first study the angular-dependence of $w_{0j\rightarrow qr}$ below.

\begin{figure*}
\includegraphics{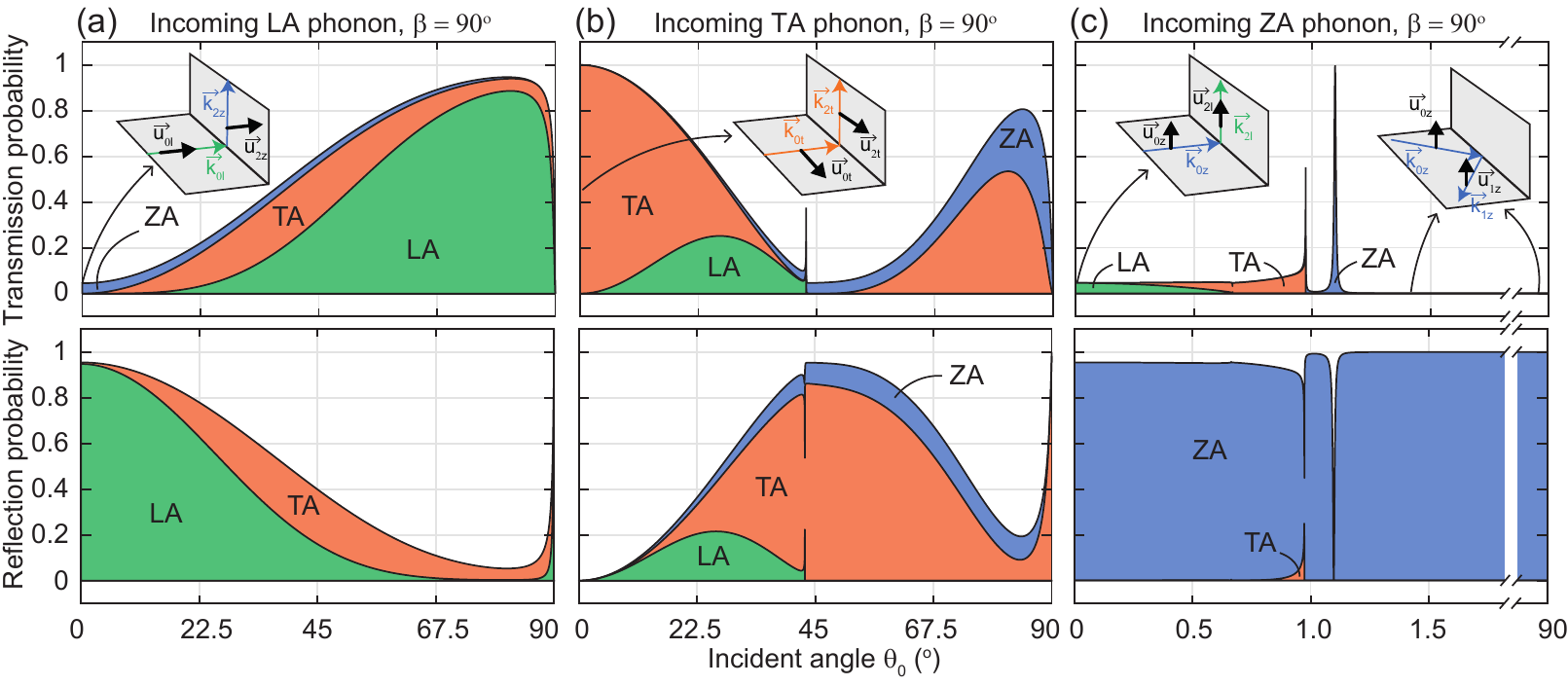}
\caption{Transmission and reflection probabilities ${w}_{ij \rightarrow qr}(\theta_0)$ as function of incident angle $\theta_0$ for (a)  LA, (b) TA and (c) ZA phonons. The insets show a sketch of the incident and transmitted phonons when $\theta_0 = 0^{\circ}$, and the additional inset in panel (c) shows the total internal reflection of the ZA phonons. Note the different x-axis in the case of Fig. (c), zooming in in the low-angle behavior.  \label{fig:GPMkink}}
\end{figure*}
\subsection{Transmission probabilities as function of incident angle for $\beta = 90^{\circ}$}
Figure \ref{fig:GPMkink} shows the angle-dependent transmission coefficients $w_{0j\rightarrow qr}(\theta_0)$ of all the three phonon modes on a graphene membrane with a pretension of $\overline{n}_{1,xx}= \overline{n}_{1,yy} = \overline{n}_{2,xx} = \overline{n}_{2,yy} =  0.03$ N/m (based on estimates from Ref.  \onlinecite{PhysRevB.96.165421}), $\overline{n}_{1,xy} = \overline{n}_{2,xy} = 0$ N/m and $\beta = 90^{\circ}$. The Lame parameters $\lambda = 15.55$ J/m$^2$ and $\mu = 103.89$ J/m$^2$ are taken from the literature \cite{atalaya2008continuum}.

The transmission of incident LA phonons is mostly affected with respect to $\beta = 0^{\circ}$ at small incident angles. This is because when $\theta_{0j} = 0^{\circ}$, $\vec{u}_{0l} \parallel \vec{u_{2z}}$, as shown in the inset of Fig. \ref{fig:GPMkink}(a). The continuity of deflection then enforces that LA phonons can only transmit into ZA phonons, which are significantly mismatched in propagation speed $c_{ij}$ ($c_{il} = \sqrt{(\lambda_i+2\mu_i)/\rho h} = 17.0$ km/s, $c_{it} = \sqrt{\mu/\rho h} = 11.6$ km/s and $c_{iz} = \sqrt{\overline{n}/\rho h} = 0.2 $ km/s, where $h = 0.335$ nm is the thickness of graphene). Using acoustic impedance mismatch theory \cite{kinsler2000fundamentals}, we obtain a transmission coefficient of $4 c_{2z} c_{1l}/(c_{2z}+c_{1l})^2 = 0.046$, matching the value obtained by the model for $\theta_{0} = 0^{\circ}$.  At larger incident angles, efficient transmission into LA and TA phonons becomes possible, raising the total transmission coefficient. 

As shown in Fig. \ref{fig:GPMkink}(b), incident TA phonons can fully transmit at small incident angles. This can also be understood from the continuity of displacement: since the amplitudes $\vec{u}_{0t} \parallel \vec{u_{2t}}$ (see inset in Fig. \ref{fig:GPMkink}(b)), incident TA phonons with $\theta_0 = 0$ can only transmit at TA phonons, meaning that there is no change in propagation speed and acoustic impedance. At an incident angle $\theta_0 = 43^{\circ}$ a sharp feature is observed.  This corresponds to the critical angle $\theta^* = \arcsin{c_{it}/c_{il}}$, where from Eq. \ref{eq:snellslaw} the angle of refraction into LA phonons would exceed 90$^{\circ}$, meaning that TA phonons can no longer be transmitted or reflected into LA phonons. 

The incoming ZA phonons in Fig. \ref{fig:GPMkink}(c) (note the horizontal axis scale) show a remarkably low transmission, due to the large propagation speed differences between in-plane and out-of-plane phonons. At very small incident angles, at an incoming angle $\theta_{0z} = 0^{\circ}$: $\vec{u}_{0z} \parallel \vec{u}_{2l}$. Since the change in acoustic impedance is the same as in the case for an incoming LA phonon at $\theta_{0l} = 0^{\circ}$, the transmission probability (0.046) is equal.  The low speed of the flexural phonons compared to the in-plane phonons results in small critical angles, the largest being $\theta^* = \arcsin{c_{iz}/c_{it}} = 0.99^{\circ}$. Above this angle, the flexural phonons can no longer reflect or transmit as LA or TA phonons, and ZA phonons are generally not transmitted. Due to this, the integrated transmission coefficient of ZA phonons is 3 orders of magnitude smaller than those of the in-plane phonons. A striking phenomenon is the transmission peak near $\theta_0 = 1.1^{\circ}$, which emerges due to a resonant excitation of waves residing at the interface. This effect resembles the formation of Rayleigh waves on the surface of the solid material interfacing with a liquid \cite{peterson1973kapitza}. Furthermore, similar interface waves have been observed between two graphene domains in semi-molecular dynamics simulations \cite{ghaffari2018modal}.

\begin{figure}[t!]
\includegraphics{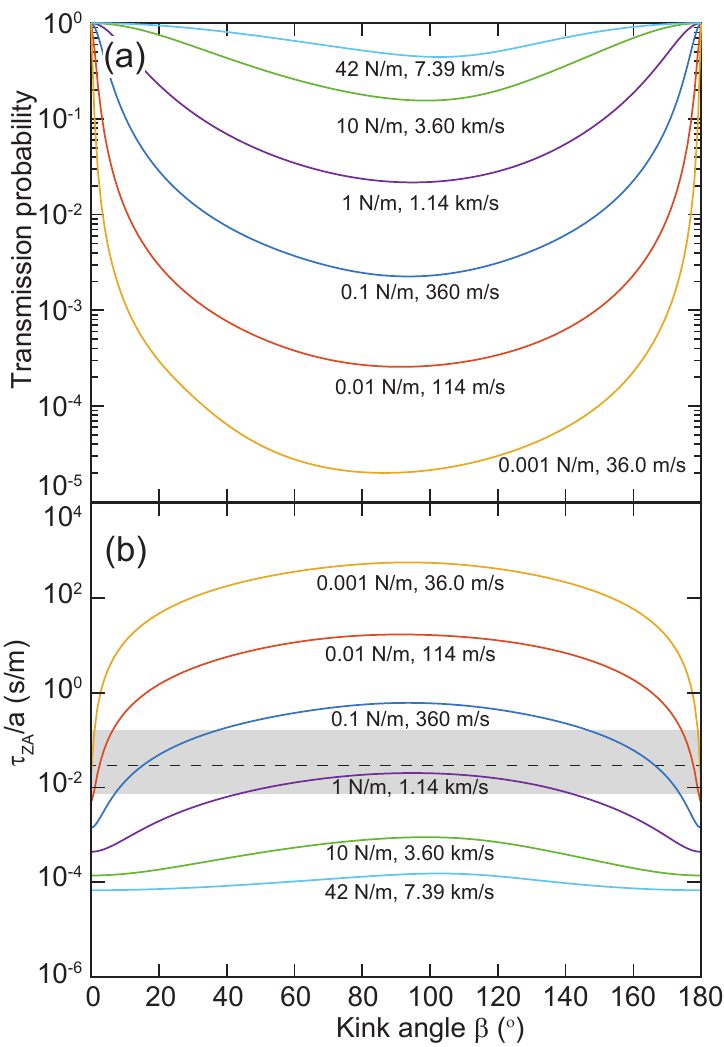}
\caption{(a) Fraction of transmitted flexural phonons $\sum_r \bar{w}_{1z \rightarrow 2r}$ for different values of the pretension as a function of kink angle $\beta$. (b) Time constant attributed to the flexural acoustic phonons $\tau_{\mathrm{ZA}}$ divided by drum radius $a$ as a function of kink angle $\beta$. The pretension is varied to show the effect of phonon propagation speed on the time constant. The gray area indicates the experimental range from Ref. \onlinecite{noncontinuum} and the dashed line the experimental mean. \label{fig:taubeta}}
\end{figure}
\section{Two-temperature model}\label{sec:tauZA}
The goal of this section is to demonstrate that the presented model is in line with the large values of thermal time constants found in Ref. \onlinecite{PhysRevB.96.165421} and the observation of the opposing thermal expansion forces in Ref. \onlinecite{noncontinuum}. 
We analyze the situation where a (optothermal) heat flux is incident at the center of a circular drum. In the case of local thermal equilibrium (where all the acoustic phonon modes have the same temperature), the boundary scattering effect presented above cannot account for the experimental observations, due to the high transmission coefficients of the in-plane phonons (see Supplemental information S2 \cite{supplemental}).  Therefore, we construct a two-temperature model to describe heat transport through suspended graphene, where the in-plane LA and TA phonons are assumed to be at a different temperature than the out-of-plane flexural ZA phonons.  It is assumed that the heat generates only in-plane acoustic phonons due to selective electron-phonon coupling \cite{singh2011spectral,vallabhaneni2016reliability}, which propagate outward from the center. Conversion between in-plane and out-of-plane phonon modes on the suspended part of the drum is neglected due to their weak mutual interactions \cite{lindsay2010flexural}. At the edge of the drum the phonons are transmitted and reflected by the kink in graphene. Due to this reflection a conversion between different phonon modes can occur, that can be analyzed by the theory from the previous section.

Thus we can determine the time-dependent internal energies of different phonon modes on the suspended part of the graphene drum. Transmitted phonons are lost, but ZA phonons can be reflected multiple times at the edge of the graphene which leads to a significantly larger value of $\tau_{\mathrm{ZA}}$, as found in experiments. 
Due to different transmission coefficients for ZA and in-plane phonons, large differences in the phonon densities, and related phonon bath temperatures of the different phonon modes, can occur. Due to this, local thermal equilibrium is violated, similar to recent predictions of \citeauthor{vallabhaneni2016reliability}\cite{vallabhaneni2016reliability}. To model this, we construct a similar two temperature model where scattering between in-plane and out-of-plane phonon modes is neglected. Instead of this, the phonon conversions at the kink are taken into account. 

To simplify the problem we note that, according to Fig. \ref{fig:GPMkink}, the in-plane phonons have a high probability of crossing the kink at the edge of the suspended graphene, and therefore experience a low thermal interface resistance. The flexural ZA phonons, on the other hand, are confined to the drum due to total internal reflection and therefore experience a large thermal interface resistance, making them responsible for the long thermal time constants $\tau_{\mathrm{ZA}}$ observed in experiments. To predict the long thermal time constant $\tau_{\mathrm{ZA}}$, this means that analysis can be simplified by initially focusing on the flexural phonons alone and explain the value of the thermal time constants observed in Ref.  \onlinecite{PhysRevB.96.165421} (subsection \ref{subsec:ZA}). After this, the model will be expanded to also include the flow of heat attributed to the in-plane acoustic phonons, to explain the opposing thermal expansion forces in Ref. \onlinecite{noncontinuum} (subsections \ref{subsec:C2C1} -- \ref{sec:distribution}). The final subsection \ref{subsec:inplane} estimates the value of the thermal time constant of the in-plane acoustic phonons, to verify that it is much shorter than that of the flexural acoustic phonons.

\subsection{Time constant for flexural phonons}\label{subsec:ZA}
In this section, we study a simplified model that predicts the time constant $\tau_{\mathrm{ZA}}$, that is compared to experimental values \cite{PhysRevB.96.165421,noncontinuum} of the time constant. This comparision allows us to estimate the average pre-tension $\bar{n}$ in the membrane, which will be used in the following subsection. Assuming the environmental temperature is higher than the Debye temperature for ZA phonons, expressions for the heat capacity $\mathcal{C}_{\mathrm{ZA}}$ and thermal resistance $\mathcal{R}_{\mathrm{ZA}} $ for a circular membrane were derived in Ref. \onlinecite{PhysRevB.96.165421}:
\begin{equation}
\mathcal{R}_{\mathrm{ZA}} = \frac{1}{G_{B,z} h 2 \pi a}=  \frac{ A_{uc}}{2 \pi a k_B \sum_r \bar{w}_{1z \rightarrow 2r}  c_{\mathrm{ZA}}},
\end{equation}
\begin{equation}
\mathcal{C}_{\mathrm{ZA}}= c_{p,z} \rho h \pi a^2 = \frac{k_B \pi a^2}{A_{uc}},
\end{equation}
where $G_{B,z}$ is the thermal boundary conductance of the ZA phonons, $h$ the thickness of graphene, $a$ the drum radius, $k_B$ the Boltzmann constant, $c_{\mathrm{ZA}}$ the propagation speed of ZA phonons and $A_{uc}$ the unit cell area of graphene. 
For a circular membrane, the flexural phonon time constant $\tau_{\mathrm{ZA}} = \mathcal{R}_{\mathrm{ZA}} \mathcal{C}_{\mathrm{ZA}}$ is described by the equation:
\begin{equation}\label{eq:tauZA}
 \tau_{\mathrm{ZA}} = \frac{a}{2 \sum_r \bar{w}_{1z \rightarrow 2r}  c_{\mathrm{ZA}} },
\end{equation}

Figure \ref{fig:taubeta} shows the transmission coefficient and time constant $\tau_{\mathrm{ZA}}$ as a function of kink angle $\beta$ and for different values of the average pretension $\bar{n}$. Since the phonon velocities on the supported and suspended graphene are equal by assumption, the transmission coefficient of the ZA phonons is equal to 1 when the kink angle is 0 or 180 degrees. The transmission coefficient already changes dramatically for small kink angles. The transmission coefficient is minimal for a kink of 90 degrees. 

We compare the model to the experimental values of $\tau/a$ found in related works \cite{PhysRevB.96.165421,noncontinuum}. In Fig. \ref{fig:taubeta}(b) the grey area indicates the highest and lowest observed value of $\tau/a$ and the dashed line indicates the mean value $\overline{\tau/a} = 0.029$~s/m. Assuming sidewall adhesion with a kink angle of 90 degrees we estimate the phonon speed to be 1.0~km/s on average, corresponding to a tension of $\sim$0.8~N/m. This value is reasonable compared to pre-tension values obtained in literature \cite{lee2008measurement}, and we will use this value in the following subsections.  

\subsection{Model for opposing thermal expansion forces}\label{subsec:C2C1}
In this subsection, we calculate the ratio between the opposing thermal expansion forces in the steady-state regime, which are found in experiments in Ref. \onlinecite{noncontinuum}. As explained above and in the Supplemental information S2 \cite{supplemental}, we expect the in-plane and flexural acoustic phonons to be at different temperatures and therefore require a two-temperature model to describe heat transport in the suspended graphene. To do this, we assume that the LA and TA phonons are always in local thermal equilibrium with each other. This is supported by the results of \citeauthor{vallabhaneni2016reliability} \cite{vallabhaneni2016reliability} who also analyzed suspended graphene heated by a laser, and found the LA and TA phonons to be at the same temperature. 
The internal energies are related to the modal temperatures by the expression \cite{PhysRevB.96.165421}:
\begin{equation}
\begin{aligned}\label{eq:intenerergies}
U_{ij} = \frac{\zeta(3) k_B^3 T_{\mathrm{LA+TA}}^3}{\pi c_{ij}^2 \hbar^2 h} ~\mathrm{for}~ j = t, l \\
U_{ij} = \frac{k_B T_{\mathrm{ZA}}}{h A_{uc}} ~\mathrm{for}~ j = z,
\end{aligned}
\end{equation}
where $\hbar$ is the reduced Planck constant, and $\zeta(3) \approx 1.21$ Apéry's constant. Using Eq. \ref{eq:intenerergies} the internal energy of the LA phonons $U_{1l}$ is related to the internal energy of the TA phonons $U_{1t}$ by:
\begin{equation}\label{convertLATA}
U_{1t} = \frac{c_{1l}^2}{c_{1t}^2} U_{1l}.
\end{equation}
Due to selective electron-phonon coupling, the LA and TA phonon modes are also the only modes that will receive the heat flux from the laser \cite{singh2011spectral,vallabhaneni2016reliability}. For the ZA phonon bath, we assume that the heat transport is limited by the Kapitza resistance induced by the kink, as this was also used to calculate $\tau_{\mathrm{ZA}}$ in subsection \ref{subsec:ZA} above. 

Using the assumptions above, we use the heat equation in cylindrical coordinates \cite{cai2010thermal,lee2011thermal} to find the change in internal energy of the in-plane phonons $\Delta U_{1l}$:
\begin{equation}\label{heateqlaser}
\frac{\kappa_{\mathrm{LA+TA}}}{\rho c_{p,\mathrm{LA+TA}}} \frac{1}{r} \frac{\mathrm{d}}{\mathrm{d} r} \left( r \frac{\mathrm{d}\Delta U_{\mathrm{1l}}}{\mathrm{d} r} \right) + Q''' = 0,
\end{equation}
where $\kappa_{\mathrm{LA+TA}}$ is the thermal conductivity of the in-plane phonon bath, $c_{p,\mathrm{LA+TA}}$ the specific heat of the in-plane phonon bath and $Q'''$ is the volumetric heat flux of the laser. This is described by the Gaussian spatial dependence:
\begin{equation}
 Q''' = Q_0 \exp{\left(\frac{-r^2}{r_0^2}\right)},
\end{equation}
where $r_0$ is the radius of the laser spot, estimated to be $r_0 =  285$ nm. Using this spatial dependence the general solution to Eq. \ref{heateqlaser} is:
 \begin{equation}\label{eq:Ula}
 U_{\mathrm{LA}}(r) = A_1 + A_2 \mathrm{ln}(r) + A_3 \mathrm{Ei}\left( \frac{-r^2}{r_0^2}\right),
 \end{equation}
where $A_1$, $A_2$ and $A_3$ are constants to be determined and $\mathrm{Ei}$ is the exponential integral function. $A_1$, $A_2$ and $A_3$  are found by enforcing a continuous solution when $r \rightarrow 0$ and applying an energy balance at the boundary of the drum. $\Delta U_{1z}$ is modeled by assuming that the thermal interface resistance at the edge of the drum is limiting the heat transport; therefore $\Delta U_{1z}$ is uniform over the suspended drum. Since $\Delta U_{1z}$ appears in the boundary conditions, solving Eq. \ref{eq:Ula} results in solutions for $\Delta U_{1l}(r)$ and $\Delta U_{1z}$ which are presented in the Supplemental information S3 \cite{supplemental}.

The force that actuates the out-of-plane motion of the membrane is proportional to the strain in the membrane \cite{dolleman2018opto}.
To find the ratio between the thermal expansion forces, one can therefore convert the internal energies to the mechanical strain contribution from each phonon mode $\Delta \epsilon_j$ using the expression \cite{ge2016comparative}:
\begin{equation}\label{eq:thermalex}
\Delta \epsilon_j = - \frac{1}{4 K} \gamma_j U_j ,
\end{equation}
where $K = 158$ GPa the bulk modulus. The ratio between the thermal expansion forces $C_{\mathrm{LA+TA}}/C_{\mathrm{ZA}} =  (\Delta \epsilon_{\mathrm{LA}} + \Delta \epsilon_{\mathrm{TA}})/\Delta \epsilon_{\mathrm{ZA}} $ becomes:
\begin{equation}\label{eq:forceratio3}
\frac{C_{\mathrm{LA+TA}}}{C_{\mathrm{ZA}}} = \frac{\gamma_{\mathrm{LA}}\Delta \bar{U}_{1l} + \gamma_{\mathrm{TA}} \frac{c_{1l}^2}{c_{1t}^2} \Delta \bar{U}_{1l}}{\gamma_{\mathrm{ZA}} \Delta U_{1z}},
\end{equation}
where $\bar{U}_{1l}$ is the average internal energy of the LA phonons over the surface of the drum. This ratio of the forces determines the mechanical out-of-plane response of the membrane, and should therefore match the force ratio observed in experiments \cite{noncontinuum}.

Evaluation of the model requires several parameters from theory. First, the in-plane thermal conductivity $k_{\mathrm{LA+TA}}$ is required, whose value can show considerable spread in literature \cite{nika2009lattice,lindsay2010flexural,xu2014length}. Second is the mode Grüneisen parameter $\gamma_{\mathrm{ZA}}$, which is difficult to calculate at low phonon frequencies \cite{mounet2005first,schelling2003thermal,mann2017negative}. Here, we use literature values of the mode Grüneisen parameters: $\gamma_{\mathrm{LA}} = 1.06$, $\gamma_{\mathrm{TA}} = 0.40$ and $\gamma_{\mathrm{ZA}} = -4.17$ from \citeauthor{mann2017negative} \cite{mann2017negative}. Finally, the angular distribution of $\theta_{0j}$ at which phonons are indicent at the boundary is of influence. For now, we assume a uniform angular distribution, but its influence will be investigated further below. 

\subsection{Modal temperatures}
\begin{figure}
\includegraphics{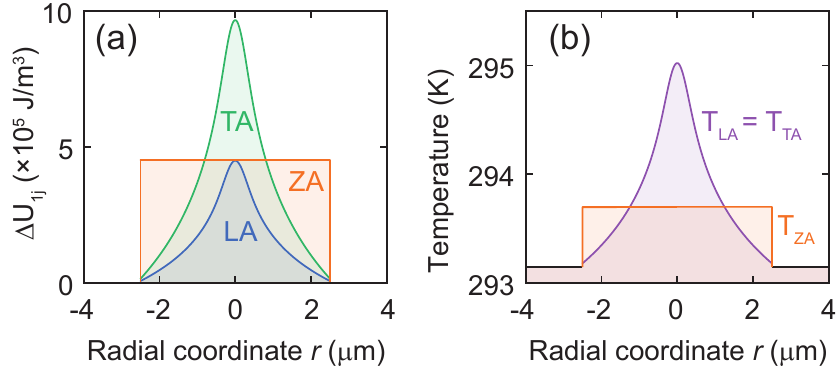}
\caption{(a) Change in internal energy and (b) modal temperature as a function of radial coordinate $r$ with in-plane thermal conductivity $\kappa_{\mathrm{LA+TA}} = 2000$ W/m K \cite{klemens1994thermal}, laser spot size $r_0 = 285$ nm, drum radius $a = 2.5$ $\mu$m and total absorbed laser power $Q_{\mathrm{laser}} = 1$ $\mu$W. \label{fig:intenergyfuncr}}
\end{figure}
First, we study the internal energy and modal temperature in the membrane as a function of position. As a starting point we take the in-plane thermal conductivity of graphite as $\kappa_{\mathrm{LA+TA}}$, which is equal to 2000 W/m K. The internal energy as a function of position $r$ is shown in Fig. \ref{fig:intenergyfuncr}(a). These values are converted to temperature in Fig. \ref{fig:intenergyfuncr}(b) by using Eq. \ref{eq:intenerergies}.
The ZA phonons show a large temperature jump due to their large Kapitza resistance. Since the rate of ZA phonon generation from the in-plane phonon bath is much higher than that of ZA phonons leaving the membrane, this phonon bath reaches relatively high internal energies, even though this bath only receives a small fraction of the total heat flux supplied to the system due to selective electron-phonon coupling. Converting the average internal energies to the force ratio (Eq. \ref{eq:forceratio3}), we find for this specific drum diameter of 5 $\mu$m and $\kappa_{\mathrm{LA+TA}} = 2000$ W/m K that $-C_{\mathrm{LA+TA}}/C_{\mathrm{ZA}} = 0.098$. Compared to experiments, the median value of $-C_{\mathrm{LA+TA}}/C_{\mathrm{ZA}} = 0.2$ for a 5 $\mu$m diameter drum, the model thus predicts values of the force ratio in the right order of magnitude.

\begin{figure}
\includegraphics{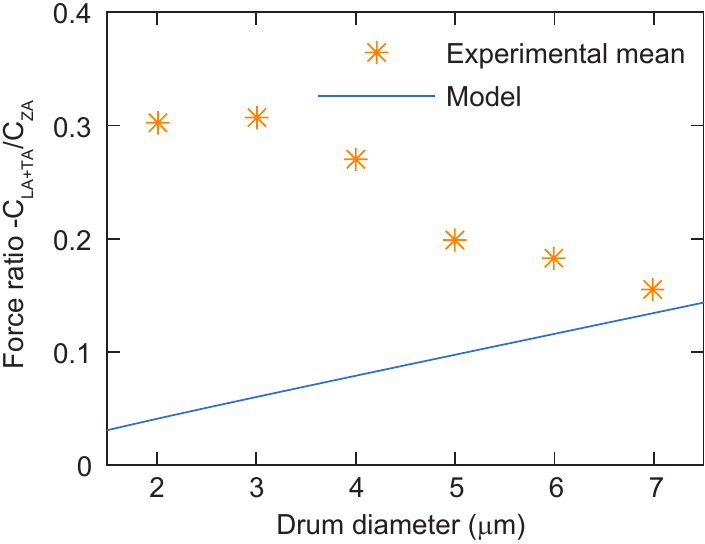}
\caption{$-C_{\mathrm{LA+TA}}/C_{\mathrm{ZA}}$ as a function of drum diameter calculated from Eq. \ref{eq:forceratio3} compared to experimental mean values from Ref. \onlinecite{noncontinuum}.  A constant value of $\kappa_{\mathrm{LA+TA}} = 2000$ W/m K and $r_0 = 285$ nm is assumed. \label{fig:diamdep}}
\end{figure}
If the ratio $-C_{\mathrm{LA+TA}}/C_{\mathrm{ZA}}$ is calculated as a function of diameter, however, the model predicts an increasing trend, while the experiments show a decreasing trend (Fig. \ref{fig:diamdep}). Likely this is due to the assumption that $\kappa_{\mathrm{LA+TA}}$ is constant as a function of diameter, while literature suggests that the effective thermal conductivity $\kappa_{\mathrm{LA+TA}}$ is length-dependent \cite{nika2009lattice,lindsay2010flexural,xu2014length}. This is because the mean free path of the in-plane phonons is not small enough compared to the drum size and, as a consequence, the phonon transport is still partly ballistic \cite{lee2015hydrodynamic,cepellotti2015phonon}. This causes boundary effects to have an important affect on the in-plane thermal conductivity $\kappa_{\mathrm{LA+TA}}$. In subsection \ref{sec:kinplane} we will investigate whether a diameter-dependent $\kappa_{\mathrm{LA+TA}}$ can account for the experimental results. 

Another consequence of the (partly) ballistic nature of the phonon transport is that the angular distribution of the phonons incident on the boundary is no longer uniform. Keeping in mind that phonons are primarily generated in the center of the drum and initially propagate radially outward, small drums have more phonons with normal incidence on the boundary. On the other hand, large drums have a more uniform distribution, as more scattering events are expected to occur between the center and the edge of the drum. As shown in Fig. \ref{fig:GPMkink}, the transmission of phonons is strongly dependent on their incident angle, and this could account for the anomalous diameter dependence of $-C_{\mathrm{LA+TA}}/C_{\mathrm{ZA}}$ observed in the experiments. Therefore, the influence of the angular distribution of incident phonons is investigated in subsection \ref{sec:distribution}.

\subsection{Influence of the in-plane thermal conductivity}\label{sec:kinplane}
\begin{figure}
\includegraphics{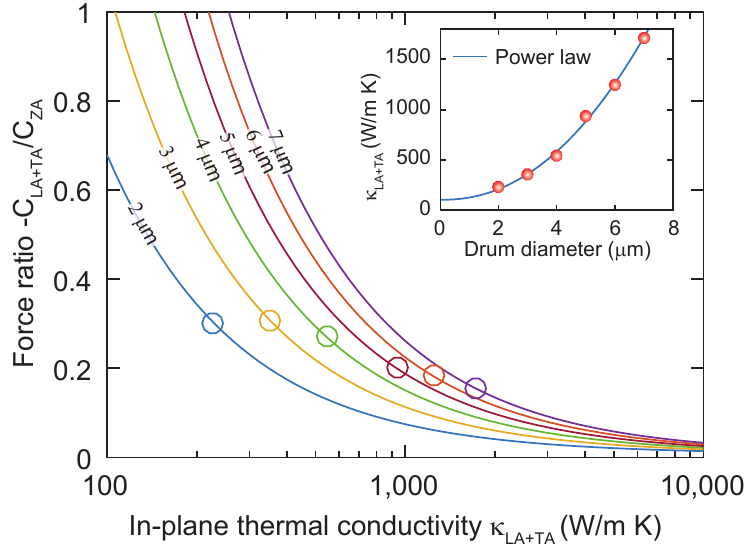}
\caption{Dependence of $-C_{\mathrm{LA+TA}}/C_{\mathrm{ZA}}$ on the thermal conductivity of the in-plane phonons plotted for different drum diameters, using the model in section \ref{subsec:C2C1}. The circles represent the experimental mean from Ref. \onlinecite{noncontinuum}. The inset shows the extracted in-plane thermal conductivity as a function of drum diameter based on the experimental mean of $-C_{\mathrm{LA+TA}}/C_{\mathrm{ZA}}$, with a power law ($\kappa_{\mathrm{LA+TA}} = c_0+c_1(2a)^p$) fit to the data. \label{fig:thermalcond}}
\end{figure}
To explain the diameter dependence of the ratio $-C_{\mathrm{LA+TA}}/C_{\mathrm{ZA}}$ in Ref. \onlinecite{noncontinuum}, we first study the effect of the thermal conductivity of the in-plane phonons $\kappa_{\mathrm{LA+TA}}$. Figure \ref{fig:thermalcond} shows the calculated ratio $-C_{\mathrm{LA+TA}}/C_{\mathrm{ZA}}$ as a function of $\kappa_{\mathrm{LA+TA}}$ for different drum diameters. As the thermal conductivity of the in-plane phonons increases, the ratio $-C_{\mathrm{LA+TA}}/C_{\mathrm{ZA}}$ decreases. This is because the in-plane phonons reach a lower temperature, which reduces the amplitude $C_{\mathrm{LA+TA}}$. Using the experimental mean of $-C_{\mathrm{LA+TA}}/C_{\mathrm{ZA}}$, the in-plane thermal conductivity needed to match theory and experiment can be extracted as shown in the inset in Fig. \ref{fig:thermalcond}. A strong increase in thermal conductivity is observed as the drum diameter increases. An increase of in-plane thermal conductivity with increasing diameter has been reported in various works \cite{nika2009lattice,lindsay2010flexural,xu2014length}. However, if we fit a power law to $\kappa_{\mathrm{LA+TA}} = c_0+c_1(2a)^p$ (see inset of Fig. \ref{fig:thermalcond}), we find an exponent $p = 2$, while in literature $p\leq 0.5$ is reported on suspended graphene with similar dimensions \cite{nika2009lattice,lindsay2010flexural}. The relative increase found in Fig. \ref{fig:thermalcond} is thus much stronger than reported in the literature. This considerable disagreement suggests that other effects should be taken into consideration to explain the diameter dependence of $-C_{\mathrm{LA+TA}}/C_{\mathrm{ZA}}$.

\subsection{Influence of angular phonon distribution}\label{sec:distribution}
\begin{figure}
\includegraphics{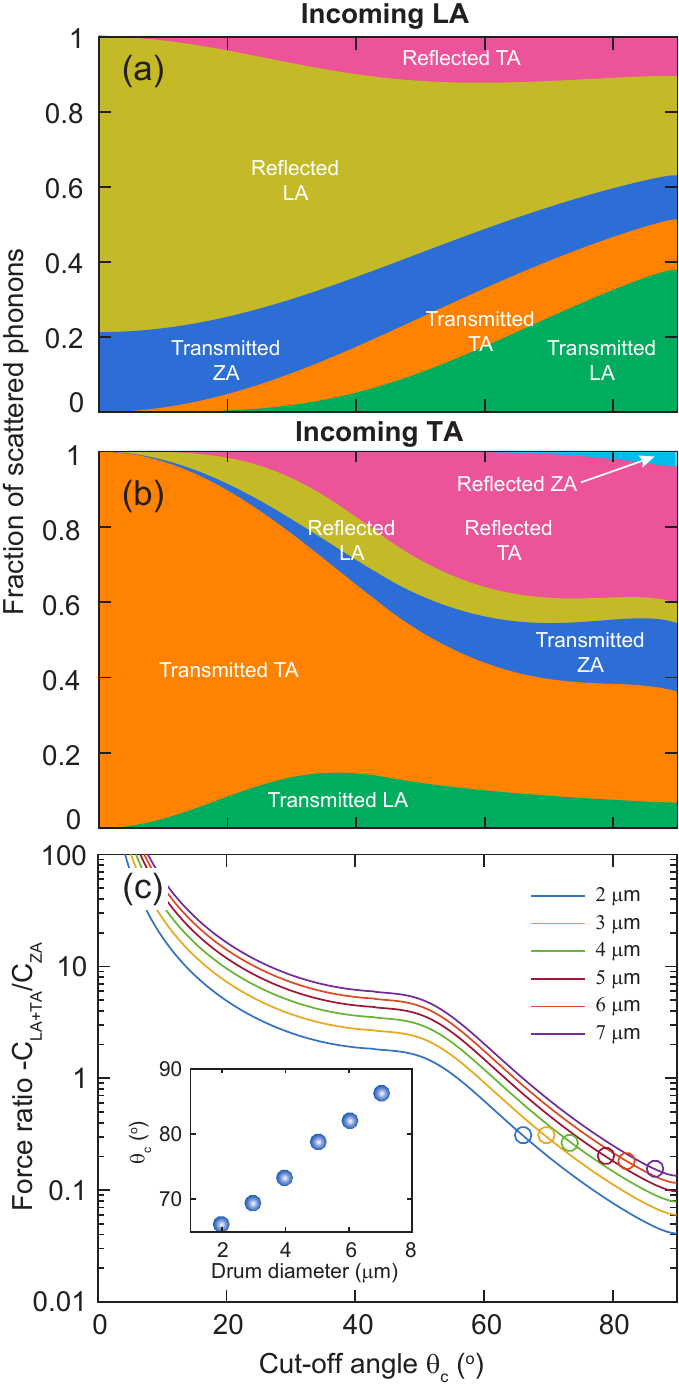}
\caption{Dependence on the angular distribution of the incoming phonons, assuming $\kappa_{\mathrm{LA+TA}} = 2000$ W/m K, using the model in section \ref{subsec:C2C1} with transmission coefficients adapted to the nonuniform angular phonon distribution. (a) Transmission and reflection probabilities for incoming longitudinal acoustic phonons as a function of cut-off angle $\theta_c$ for a pre-tension of 0.8 N/m. (b) Transmission and reflection probabilities for incoming transverse acoustic phonons as a function of cut-off angle $\theta_c$ for a pre-tension of 0.8 N/m. (c) Ratio $-C_{\mathrm{LA+TA}}/C_{\mathrm{ZA}}$ as a function of cut-off angle $\theta_c$ for different drum diameters. The circles represent the experimental mean from Ref. \cite{noncontinuum}. The inset shows the extracted cut-off angle $\theta_c$ based on the experimental mean values.   \label{fig:angledependence}}
\end{figure}
So far we have assumed the incoming angular distribution of the phonons to be uniform. However, since the mean free path of the phonons is not necessarily much shorter than the size of the suspended membrane \cite{lee2015hydrodynamic,cepellotti2015phonon,singh2011spectral}, a non-uniform angular distribution is expected. Therefore, in this section, we alter the incoming phonon distribution with a distribution function $f(\theta_0)$, to analyze the influence of a nonuniform angular distribution on the ratio $-C_{\mathrm{LA+TA}}/C_{\mathrm{ZA}}$, using the model in section \ref{subsec:C2C1}.
We adapt the integration of the transmission probabilities to include $f(\theta_0)$, which is the normalized incident phonon distribution:
\begin{equation}
\bar{w}_{ij\rightarrow qr} = \frac{2}{\pi}\int\limits_0^{\pi/2} f(\theta_0) {w}_{ij\rightarrow qr}(\theta_0) \mathrm{d} \theta_0.
\end{equation}
We simplify the analysis by only taking into account variations in $f(\theta_0)$ for the LA and TA phonons, since this is the bath where phonons are primarily generated. The heat flows consecutively into the ZA phonons and this phonon bath experiences many collisions at the boundary, therefore this angular distribution is assumed to be uniform. The incoming phonon distribution of the LA and TA phonons is altered by the following step function:
\begin{equation}
f(\theta_0) =
  \begin{cases}
    \pi  /2  \theta_c     & \quad \text{if } \theta_0 \leq \theta_c \\
    0 & \quad \text{if } \theta_0 > \theta_c
  \end{cases}
\end{equation}
where $\theta_c$ is a cut-off angle above which there are no incident phonons on the boundary. For simplicity, it is assumed that $\theta_c$ is equal for the in-plane and out-of-plane phonons.

Figure \ref{fig:angledependence}(a) shows the integrated transmission probabilities for the incoming LA phonons $\bar{w}_{1l\rightarrow qr}$ as a function of the cut-off angle and Fig. \ref{fig:angledependence}(b) shows $\bar{w}_{1t\rightarrow qr}$. The resulting value of $-C_{\mathrm{LA+TA}}/C_{\mathrm{ZA}}$ as a function of $\theta_c$ is shown in Fig. \ref{fig:angledependence}(c), for different drum diameters. To construct this figure, a value of $\kappa_{\mathrm{LA+TA}} = 2000$ W/m K is assumed for all the drum diameters. The most important process that alters the value of $-C_{\mathrm{LA+TA}}/C_{\mathrm{ZA}}$ is the reflection of TA phonons into ZA phonons, as this governs the temperature of the ZA phonon bath, and this can only occur at incident angles $\theta_0 \neq 0$. Therefore at low incident angles in Fig. \ref{fig:angledependence}, $-C_{\mathrm{LA+TA}}/C_{\mathrm{ZA}}$ becomes very large because the ZA phonons receive no heat directly from the laser, and therefore reach a low temperature compared to the in-plane phonons. At angles above $\theta_c \approx 45$ degrees the reflection of TA phonons into ZA phonons becomes significant (Fig. \ref{fig:angledependence}(b)), resulting in a sharp decrease of $-C_{\mathrm{LA+TA}}/C_{\mathrm{ZA}}$  (Fig. \ref{fig:angledependence}(c)). 

Using the experimental values of $-C_{\mathrm{LA+TA}}/C_{\mathrm{ZA}}$ from Ref. \onlinecite{noncontinuum}, a diameter dependent $\theta_c$ can be extracted as shown in the inset of Fig. \ref{fig:angledependence}(c). Values of $\theta_c$ close to 90 degrees suggest the angular distribution is close to uniform, and the LA and TA phonons are closer to the fully diffusive regime rather than the fully ballistic regime. A monotonically increasing $\theta_c$ is obtained with increasing drum size, as expected due to the increased amount of collisions experienced by the phonons as the distance between the laser spot and the boundary becomes larger, increasing the uniformity of the incoming angular phonon distribution.  This scenario is therefore a reasonable explanation to the experimentally observed diameter dependence of $-C_{\mathrm{LA+TA}}/C_{\mathrm{ZA}}$. 

\subsection{Time constant of the in-plane phonons}\label{subsec:inplane}
In Ref. \onlinecite{noncontinuum}, it is argued that the thermal time constant of the in-plane phonons must be much smaller than that of the flexural phonons. Since it is complicated to solve the time-dependence of the heat flow in the entire system, we estimate $\tau_{\mathrm{LA+TA}}$ using a simple model \cite{PhysRevB.96.165421,aubin2004radio,bunch2008mechanical} based on the solution of the heat equation and by assuming the interfacial thermal resistance of the in-plane phonons to be small:
\begin{equation}
\tau_{\mathrm{LA+TA}} \approx \frac{a^2 \rho c_{p,\mathrm{LA+TA}}}{2 \kappa_{\mathrm{LA+TA}}}.
\end{equation}
Using the values of $\kappa_{\mathrm{LA+TA}}$ from Fig. \ref{fig:thermalcond}, we find $\tau_{\mathrm{LA+TA}} \approx 2$ ns. This is indeed much smaller than the observation limit in Ref. \onlinecite{noncontinuum}. The model presented in this work thus supports the notion in Ref. \onlinecite{noncontinuum} that $\tau_{\mathrm{LA+TA}} \ll \tau_{\mathrm{ZA}}$, because typically $\tau_{\mathrm{ZA}}$ is found in a range between 25 and 250 ns. 

\section{Discussion}\label{sec:disc}
In future work, our model could be improved by taking into account the finite radius of the kink due to the bending rigidity of 2D materials \cite{lindahl2012determination,ghaffari2018modal}, which will provide a more accurate picture for the reflection and transmission of phonons with short wavelengths. 
Furthermore, coupling to the substrate could be included as an additional pathway to transmit phonons to the heat sink. 
Moreover, solutions of the full Boltzmann-Peierls equation for phonon transport in graphene \cite{peraud2014monte,lindsay2014phonon,landon2014deviational} can be useful to take into account the non-uniform angular distribution in a more accurate manner.
Finally, the model could be improved by including the anharmonic conversion processes between in-plane acoustic phonons and flexural acoustic phonon on the suspended drum \cite{vallabhaneni2016reliability,lindsay2010flexural}.

Future experiments to test our model in more detail could focus on the dependence of  $\tau_{\mathrm{ZA}}$ and $-C_{\mathrm{LA+TA}}/C_{\mathrm{ZA}}$ on the tension and the kink angle $\beta$. For example, MEMS devices could be used to strain a suspended sheet of graphene \cite{Goldsche_2018}, which should induce significant changes in $\tau_{\mathrm{ZA}}$. Also inflated graphene blisters, such as studied by \citeauthor{bunch2012adhesion}\cite{bunch2012adhesion}, provide a way to introduce large changes in the kink angle $\beta$. These studies of $\beta$ and strain could also shed more light on the large device-to-device variations observed in the experimental value of $\tau_{\mathrm{ZA}}$ \cite{PhysRevB.96.165421,noncontinuum}. Although on larger length scales experimental techniques are available \cite{wolfe2005imaging} to study the angular dependence of phonon transmission as in Fig. \ref{fig:GPMkink}, these need to be scaled down further in order to be applicable for 2D materials. If this can be overcome, it would be particularly interesting to verify the transmission peak for ZA phonons that is observed near 1.1$^{\circ}$ in Fig. \ref{fig:GPMkink}. Since Raman spectroscopy techniques to measure heat transport are mostly sensitive to the temperature of the in-plane phonon bath, they can also be useful to refine the modeling of the in-plane phonons.

\section{Conclusion}\label{sec:concl}
We analyze the situation where a (optothermal) heat flux is incident at the center of a circular graphene drum. It is assumed that the heat generates only in-plane acoustic phonons, due to selective electron-phonon scattering, that propagate outward. Due to the weak interactions between in-plane and flexural phonons, only at the edge of the drum conversion between the phonon modes can occur. Here, the phonons are transmitted and reflected by a kink in graphene that is formed by sidewall adhesion. Due to the large difference between the transmission coefficients for ZA and in-plane phonons, large differences in the acoustic phonon bath temperatures can occur. This creates a situation where the local thermal equilibrium assumption is not valid anymore on the drum. In particular, flexural phonons show a low transmission probability because their propagation speed is much lower than the in-plane phonons, which leads to a large thermal interface resistance at the edge of the drum. This resistance results in large values of the thermal time constant $\tau_{\mathrm{ZA}}$, which is in line with experimental observations. Furthermore, the different phonon temperatures lead to two distinct thermal expansion forces in suspended graphene, that oppose each other. The model predicts the ratio of the amplitudes of these forces in the correct order of magnitude observed in experiments, and shows that size dependence of this ratio can emerge due to ballistic effects in the phonon transport.

\begin{acknowledgements}
The authors thank D.R. Ladiges and J.E. Sader for fruitful discussions. This work is part of the research programme Integrated Graphene Pressure Sensors (IGPS) with project number 13307 which is financed by the Netherlands Organisation for Scientific Research (NWO).
The research leading to these results also received funding from the European Union's Horizon 2020 research and innovation programme under grant agreement No 785219 Graphene Flagship. 
\end{acknowledgements}

\newpage
\begin{widetext}
\section*{Supplemental information}
\section*{S1: Expressions for the continuity of deflection and stress} 
Here we present the complete expressions for the continuity of deflection and the continuity of stress from section II of the main text, which are used to calculate the amplitudes $\vec{u}_{ij}$ of the reflected and refracted waves. 
First, we choose the origin $t = x = y = z = 0$ at the position where the incoming phonon hits the kink. 
Then, each amplitude $\vec{u}_{2j}$ on domain 2 is projected onto the axes of domain 1, taking into account the kink angle $\beta$, and the propagation direction with respect to the normal $\theta_{ij}$. The continuity of deflection results in 3 expressions $\Sigma \vec{q}_{1j} = \Sigma \vec{q}_{2j}$ for each of the axes of domain 1:
\begin{itemize}
\item In the $\hat{\textbf{x}}$ direction:
\begin{eqnarray}
\vec{u}_{0l}\sin{\theta_{0l}} + \vec{u}_{0t}\cos{\theta_{0t}} + \vec{u}_{1l}\sin{\theta_{1l}} + 
\vec{u}_{1t} \cos{\theta_{1t}}  - \vec{u}_{2l} \sin{\theta_{2l}} - \vec{u}_{2t} \cos{\theta_{2t}} =0 \label{eq:contdefl1}
\end{eqnarray}
\item In the $\hat{\textbf{y}}$ direction:
\begin{eqnarray}
\vec{u}_{0l} \cos{\theta_{0l}} - \vec{u}_{0t} \sin{\theta_{0t}}  - \vec{u}_{1l} \cos{\theta_{1l}} + \vec{u}_{1t} \sin{\theta_{1t}} - \vec{u}_{2l}\cos{\theta_{2l}} \cos{\beta} + \nonumber \\ \vec{u}_{2t} \sin{\theta_{2t}} \cos{\beta}  + \vec{u}_{2z} \sin{\beta}= 0 \label{eq:contdefl2}
\end{eqnarray}
\item In the $\hat{\textbf{z}}$ direction:
\begin{equation}\label{eq:contdefl3}
\vec{u}_{0z} + \vec{u}_{1z} - \vec{u}_{2l} \cos{\theta_{2l}} \sin{\beta} + \vec{u}_{2t} \sin{\theta_{2t}} \sin{\beta} - \vec{u}_{2z}  \cos{\beta} =0.
\end{equation}
\end{itemize}

\subsection*{Continuity of stress}
Here we write out the full continuity of stress equations. First, taking the tension components and projecting these into the axes of domain 1, we obtain the equations:
\begin{itemize}
\item in $\hat{\textbf{x}}$ direction:
\begin{equation}
\sum\limits_{j}^{~}  \delta {n}_{1j,xy} = \sum\limits_{j}^{~}  \delta {n}_{2j,x'y'},
\end{equation}
\item in $\hat{\textbf{y}}$ direction:
\begin{equation}
\begin{aligned}
\sum\limits_{j}^{~}  \delta {n}_{1j,yy} = \overline{{n}}_{2,y'y'} \frac{\mathrm{d} \vec{q}_{2z}}{\mathrm{d} y'} \hat{\mathbf{y}}' \sin{\beta} + \sum\limits_{j}^{~}  \delta {n}_{2j, y' y'} \cos{\beta} + \overline{{n}}_{2,x'y'} \frac{\mathrm{d} \vec{q}_{2z}}{\mathrm{d} x'} \hat{\mathbf{x}}' \sin{\beta},
\end{aligned}
\end{equation}
\item in $\hat{\textbf{z}}$ direction:
\begin{equation}
\begin{aligned}
\overline{{n}}_{1,yy} \frac{\mathrm{d} \vec{q}_{1z}}{\mathrm{d} y} \hat{\mathbf{y}} + \overline{{n}}_{1,xy} \frac{\mathrm{d} \vec{q}_{1z}}{\mathrm{d} x} \hat{\mathbf{x}} = \sum\limits_{j}^{~}  \delta {n}_{2j, y' y'} \sin{\beta}  + \overline{{n}}_{2,y'y'} \frac{\mathrm{d} \vec{q}_{2z}}{\mathrm{d} y'} \hat{\mathbf{y}}' \cos{\beta} + \overline{{n}}_{2 x' y'} \frac{\mathrm{d} \vec{q}_{2z}}{\mathrm{d} x'} \hat{\mathbf{x}}' \cos{\beta},
\end{aligned}
\end{equation}
\end{itemize}
Next, we  relate the stress components to the deflections of the waves $\vec{q}_{ij}$. Taking $\vec{q}_{ij} (x,y,t) = \vec{u}_{ij}\cos(\omega t + {k}_x x + {k}_y y)$, as shown in the main text, and substituting these in the three continuity of stress equations, we obtain three expressions:

\begin{itemize}
\item  In the $\hat{\textbf{x}}$ direction:
\begin{eqnarray}
2 \mu_1 \frac{\vec{u}_{0l} \omega \sin{\theta_{0l}} \cos{\theta_{0l}}}{c_{0l}} + \mu_1 \frac{\vec{u}_{0t} \omega  \cos{2\theta_{0t}}}{c_{0t}} - 2 \mu_1 \frac{\vec{u}_{1l} \omega \cos{\theta_{1l}} \sin{\theta_{1l}}}{c_{1l}}-  \nonumber \\  \mu_1 \frac{\vec{u}_{1t} \omega \cos{2\theta_{1t}}}{c_{1t}}- 2 \mu_2 \frac{\vec{u}_{2l} \omega \sin{\theta_{2l}} \cos{\theta_{2l}}}{c_{2l}}  - \mu_2 \frac{\vec{u}_{2t} \omega  \cos{2\theta_{2t}}}{c_{2t}} =0
\end{eqnarray}
\item in the $\hat{\textbf{y}}$ direction:
\begin{eqnarray}
 2\mu_1 \frac{\vec{u}_{0l} \omega \cos^2 {\theta_{0l}}}{c_{0l}} + \lambda_1 \frac{\vec{u}_{0l}\omega}{c_{0l}} - 2\mu_1 \frac{\vec{u}_{0t} \omega \cos {\theta_{0t}}\sin{\theta_{0t}}}{c_{0t}}  +  2\mu_1 \frac{\vec{u}_{1l} \omega \cos^2{\theta_{0l}}}{c_{1l}} +  \lambda_1 \frac{\vec{u}_{1l} \omega}{c_{1l}} \nonumber \\ - 2\mu_1 \frac{\vec{u}_{1t} \omega \cos{\theta_{1t}} \sin{\theta_{1t}}}{c_{1t}}+ \overline{{n}}_{2,y'y'} \vec{u}_{2z} \frac{\omega \cos{\theta_{2z}}}{c_{2z}} \sin{\beta} - 2\mu_2 \frac{\vec{u}_{2l} \omega \cos^2 {\theta_{2l}}}{c_{2l}} \cos{\beta}  -  \nonumber \\  \lambda_2 \frac{\vec{u}_{2l} \omega }{c_{2l}} \cos{\beta}  + 2\mu_2 \frac{\vec{u}_{2t} \omega \cos {\theta_{2t}}\sin{\theta_{2t}}}{c_{2t}} \cos{\beta} -\overline{{n}}_{2,x'y'}   \vec{u}_{2z} \frac{\omega \sin{\theta_{2z}}}{c_{2z}} \sin{\beta}
  = 0
\end{eqnarray}
\item in the $\hat{\textbf{z}}$ direction:
\begin{eqnarray}
  \vec{u}_{0z}\overline{{n}}_{1,yy} \frac{\omega \cos{\theta_{0z}}}{c_{0z}} - \vec{u}_{1z} \overline{{n}}_{1,yy} \frac{\omega \cos{\theta_{1z}}}{c_{1z}} + \vec{u}_{0z}  \overline{{n}}_{1,xy} \frac{\omega \sin{\theta_{0z}}}{c_{0z}} + \vec{u}_{1z} \overline{{n}}_{1,xy} \frac{\omega \sin{\theta_{1z}}}{c_{1z}}  \nonumber \\
  -  2\mu_2    \sin{\beta} \frac{\vec{u}_{2l} \omega \cos^2 {\theta_{2l}}}{c_{2l}} - \lambda_2    \sin{\beta} \frac{\vec{u}_{2l} \omega }{c_{2l}} + 2\mu_2   \sin{\beta} \frac{\vec{u}_{2t} \omega \cos {\theta_{2t}}\sin{\theta_{2t}}}{c_{2t}} - \nonumber \\ \vec{u}_{2z} \overline{{n}}_{2,y'y'} \frac{\omega \cos{\theta_{2z}}}{c_{2z}} \cos{\beta}-  \overline{{n}}_{2 x' y'} \vec{u}_{2z} \frac{\omega \sin{\theta_{2z}}}{c_{2z}}\cos{\beta} 
     =
0
\end{eqnarray}
\end{itemize}

\subsection*{Evaluating the continuity relations}
Here we write down the continuity of deformation and continuity of stress equations in matrix form, in order to facilitate the reproduction of the calculations presented in this work. The equations are expressed in the form $\textbf{A}\vec{u} = \textbf{B}\vec{u}_{0j}$ that will enable us to evaluate the transmission probabilies of phonons for a given frequency and angle assuming that $u_{0j} = 1$.
\begin{eqnarray}
\begin{bmatrix}\label{eq:matrixLA}
a_{11} & a_{12} & a_{13} & a_{14} & a_{15} & a_{16} \\
a_{21} & a_{22} & a_{23} & a_{24} & a_{25} & a_{26} \\
a_{31} & a_{32} & a_{33} & a_{34} & a_{35} & a_{36} \\
a_{41} & a_{42} & a_{43} & a_{44} & a_{45} & a_{46} \\
a_{51} & a_{52} & a_{53} & a_{54} & a_{55} & a_{56} \\
a_{61} & a_{62} & a_{63} & a_{64} & a_{65} & a_{66} 
 \end{bmatrix}
 \begin{bmatrix}
\vec{u}_{1l}\\
 \vec{u}_{1t} \\
\vec{u}_{1z} \\
\vec{u}_{2l} \\
\vec{u}_{2t} \\
\vec{u}_{2z}
 \end{bmatrix}
=
 \begin{bmatrix}
b_1\\
b_2\\
b_3\\
b_4\\
b_5\\
b_6\\
  \end{bmatrix} u_{0j} 
  \end{eqnarray}
 where the first three rows represent the continuity equations for deflections in $x,y,z$ direction, respectively, and the bottom three rows the continuity of stress in the x,y,z directions. The coefficients of the matrices are given by:
 \begin{equation}
 \begin{alignedat}{3} 
&a_{11} = \sin{\theta_{1l}}, ~ &
&a_{12} = \cos{\theta_{1t}}, ~ &
&a_{13} = 0, ~ \nonumber \\
&a_{14} = -\sin{\theta_{2l}}, ~ &
&a_{15} = - \cos{\theta_{2t}}, ~ &
&a_{16} = 0 \nonumber \\
&a_{21} = - \cos{\theta_{1l}}, ~ &
&a_{22} = \sin{\theta_{1t}}, ~ &
&a_{23} = 0, ~ \nonumber \\
&a_{24} = -\cos{\theta_{2l}} \cos{\beta}, ~&
&a_{25} =  \sin{\theta_{2t}} \cos{\beta}, ~&
&a_{26} = \cos{\theta_{2z}} \sin{\beta}, ~ \nonumber \\
&a_{31} = 0, ~ & 
&a_{32} = 0,~ &
&a_{33} = 1, ~ \nonumber \\
&a_{34} =  - \cos{\theta_{2l}} \sin{\beta}, ~&
&a_{35} =  \sin{\theta_{2t}} \sin{\beta}, ~&
&a_{36} = - \cos{\theta_{2z}}\cos{\beta}, ~ \nonumber \\
&a_{41} = - 2 \mu_1 \frac{\cos{\theta_{1l}} \sin{\theta_{1l}}}{c_{1l}}, ~&
&a_{42} =  -  \mu_1 \frac{\cos{2\theta_{1t}}}{c_{1t}}, ~&
&a_{43} = 0, ~ \nonumber \\
&a_{44} = - 2 \mu_2 \frac{\sin{\theta_{2l}} \cos{\theta_{2l}}}{c_{2l}}, ~&
&a_{45} = - \mu_2 \frac{\cos{2 \theta_{2t}}}{c_{2t}}, ~ &
&a_{46} = 0 \nonumber \\
& a_{51} = 2\mu_1\frac{\cos^2{\theta_{1l}}}{c_{1l}} + \frac{\lambda_1}{c_{1l}}, ~ &
&a_{52} = - 2 \mu_1 \frac{\cos{\theta_{1t}} \sin{\theta_{1t}}}{c_{1t}}, ~ &
&a_{53} =  0, ~ \nonumber \\
&a_{54} = - 2\mu_2 \cos{\beta} \frac{\cos^2{\theta_{2l}}}{c_{2l}} - \frac{\lambda_2}{c_{2l}}\cos{\beta}, ~ &
&a_{55} =  2\mu_2 \cos{\beta} \frac{\cos{\theta_{2t}} \sin{\theta_{2t}}}{c_{2t}}, ~ & ~ \nonumber \\
\noalign{$\displaystyle a_{56} =  \bar{n}_{2,y'y'} \sin{\beta} \frac{\cos{\theta_{2z}}}{c_{2z}} - \bar{n}_{2,x'y'} \sin{\beta} \frac{\sin{\theta_{2z}}}{c_{2z}}$}
&a_{61} = 0, ~ &
&a_{62} =0, ~ & ~\nonumber \\
&a_{63} = - \bar{n}_{1,yy}\frac{\cos{\theta_{1z}}}{c_{1z}} -  \bar{n}_{1,xy} \frac{\sin{\theta_{1z}}}{c_{1z}}, ~ &
& a_{64} = - 2 \mu_2 \sin{\beta} \frac{\cos^2{\theta_{2l}}}{c_{2l}} - \sin{\beta}\frac{\lambda_2}{c_{2l}}, ~ & ~\nonumber \\ &a_{65} =  2\mu_2 \sin{\beta} \frac{\cos{\theta_{2t}} \sin{\theta_{2t}}}{c_{2t}} ~\nonumber \\
\noalign{$ \displaystyle a_{66} = -\bar{n}_{2,y'y'} \cos{\beta} \frac{\cos{\theta_{2z}}}{c_{2z}} - \bar{n}_{2,x'y'} \cos{\beta} \frac{\sin{\theta_{2z}}}{c_{2z}} $}
\end{alignedat}
\end{equation}
For incoming LA phonon:
\begin{equation}
\begin{alignedat}{3}
& b_1 = -\sin{\theta_{0l}}, ~ &
& b_2 = -\cos{\theta_{0l}}, ~ &
& b_3 =0, ~ \nonumber \\
&b_4 = - 2\mu_1 \frac{\sin{\theta_{0l}}\cos{\theta_{0l}}}{c_{0l}}, ~ &
&b_5 = -2\mu_1 \frac{\cos^2{\theta_{0l}}}{c_{0l}} - \frac{\lambda_1}{c_{0l}}, ~ &
& b_6 = 0. \nonumber
\end{alignedat}
\end{equation}
For incoming TA phonon:
\begin{equation}
\begin{alignedat}{3}
&b_1 = - \cos{\theta_{0t}}, ~ &
&b_2 = \cos{\theta_{0t}}, ~ &
& b_3 = 0, ~ \nonumber \\
& b_4 = -\mu_1 \frac{\cos{2\theta_{0t}}}{c_{0t}}, ~ &
& b_5 = 2 \mu_1 \frac{\cos{\theta_{0t}}\sin{\theta_{0t}}}{c_{0t}}, ~ &
& b_6 =0. \nonumber
\end{alignedat}
\end{equation}
For an incoming ZA phonon:
\begin{equation}
\begin{alignedat}{3}
&b_1 = 0, ~ &
&b_2 = 0, ~ & 
&b_3 = -1, ~ \nonumber \\
&b_4 = 0, ~ &
& b_5 =0, ~ & 
& b_6 = - \bar{n}_{1,yy} \frac{\cos{\theta_{0z}}}{c_{0z}} - \bar{n}_{1,xy} \frac{\sin{\theta_{0z}}}{c_{0z}}
\end{alignedat}
\end{equation}

\section*{S2: Thermal interface resistance in the case of local thermal equilibrium}\label{sec:TIR}
 In this section, we calculate the value of the thermal boundary resistance induced by the kink in the case of local thermal equilibrium and compare this to experimental values from ref.  \onlinecite{PhysRevB.96.165421}. In ref. \onlinecite{PhysRevB.96.165421} , a model for the thermal boundary conductance $G_B$ was derived:
\begin{equation}\label{eq:gpmresistance}
\begin{aligned}
G_B=\frac{  3 \zeta(3)  k_B^3 T^2} { \pi  \hbar^2 h_g  }  \left( \frac{\Sigma_r \bar{w}_{1l\rightarrow 2r}}{c_{1l}} + \frac{\Sigma_r \bar{w}_{1t\rightarrow 2r}}{c_{1t}} +\frac{\pi \hbar^2 \Sigma_r \bar{w}_{1z \rightarrow 2r} c_{1z}}{k_B^2 T^2 \zeta(3) A_{uc}}  \right),
\end{aligned}
\end{equation} 
where $k_B$ is Boltzmann's constant, $T$ the environmental temperature, $\hbar$ the reduced Planck's constant, $h_g = 0.335$ nm the thickness of graphene, $c_{1j}$ the phonon propagation speed on the suspended drum, $A_{uc}= \num{5e-20}$ m$^2$ the area of a unit cell of graphene and $\sum_r \bar{w}_{1j \rightarrow 2r} = \bar{w}_{1j \rightarrow 2l} +\bar{w}_{1j \rightarrow 2t} + \bar{w}_{1j \rightarrow 2z}$ is the total fraction of transmission for phonon of mode $j$ incident on the boundary. Using the transmission probabilities found in Fig. 4 in the main section of the paper and integrating them over all incoming angles, we can evaluate this model and we find a thermal boundary conductance of $G_B = 5.3$ GW/(m$^2 \cdot$K). The experimentally determined value of $G_B$ in Ref. \onlinecite{PhysRevB.96.165421} is 30 MW/(m$^2 \cdot$K), which is two orders of magnitude smaller than the value from Eq. \ref{eq:gpmresistance}. 

We attribute this discrepancy to the assumption behind Eq. \ref{eq:gpmresistance} that the in-plane and out-of-plane modes interact strongly with each other and are locally at the same temperature. However, the low experimental value of $G_B$, the long thermal time constants $\tau$, and the opposing thermal expansion forces are indications that the flexural phonons are not at the same temperature as the in-plane phonons. This further motivates the use of the two-temperature model in the main part of this work.

\section*{S3: Solutions for the internal energies in the two-temperature model}
Here, we work out the solution for the two-temperature model in more detail. 
To find the boundary conditions, each phonon scattering process has to be converted into a boundary heat flux $Q_{ij\rightarrow qr}$ \cite{PhysRevB.96.165421}: 
\begin{equation}\label{eq:heatfluxboundary}
Q_{ij\rightarrow qr} = 2\pi a h_g (\Delta U_{ij}) c_{ij} \bar{w}_{ij \rightarrow qr}.
\end{equation}
To find $A_1$, $A_2$ and $A_3$, we first take the condition that the solution must be continuous as $r \rightarrow 0$. Second, by applying conservation of energy at the boundary we find the condition:
\begin{equation}
\frac{2 \pi a h_g \kappa_{\mathrm{LA+TA}}}{\rho c_{p,\mathrm{LA+TA}}} \frac{\mathrm{d} \Delta U_{1l}(r=a)}{\mathrm{d}r} = - Q_{\mathrm{laser}}
\end{equation}
where $Q_{\mathrm{laser}}$ the total heat flux supplied by the laser. Third, we take the phonon scattering at the kink into account:
\begin{equation}
D_1 \Delta U_{1l}(r=a) + D_2 \Delta U_{1z} = Q_{\mathrm{laser}}
\end{equation}
where:
\begin{equation}
\begin{aligned}
D_1 = - 2\pi a h_g ( \bar{w}_{1l\rightarrow1z} c_{1l} +\bar{w}_{1t\rightarrow1z} \frac{c_{1l}^2}{c_{1t}}\\+ \Sigma_r \bar{w}_{1l\rightarrow2r} c_{1l}+ \Sigma_r \bar{w}_{1t\rightarrow2r} \frac{c_{1l}^2}{c_{1t}} )\\
D_2 = 2 \pi a h_g (\bar{w}_{1z\rightarrow1l} + \bar{w}_{1z\rightarrow1t})c_{1z}.
\end{aligned}
\end{equation}

Taking these boundary conditions, we find for the internal energy as function of radius:
\begin{equation}
 \Delta U_{1l}(r)  = \frac{\rho c_{p,\mathrm{LA+TA}} Q_{\mathrm{laser}}}{{2\pi h_g \kappa_{\mathrm{LA+TA}}}(1- \exp{\frac{-a^2}{r_0^2}})}\left[\mathrm{ln}\left( \frac{a}{r} \right) +\frac{1}{2}\mathrm{Ei}\left( \frac{-r^2}{r_0^2}\right) - \frac{1}{2}\mathrm{Ei}\left( \frac{-a^2}{r_0^2}\right) \right]  - \frac{D_2}{D_1} \Delta U_{1z} + \frac{Q_{\mathrm{laser}}}{D_1} ,
\end{equation}
where $\mathrm{Ei}(x)$ is the exponential integral of $x$. To convert this to the mechanical response, we take the average over the drum surface:
\begin{equation}
\Delta \bar{U}_{1l}  = 
\frac{\rho c_{p,\mathrm{LA+TA}} Q_{\mathrm{laser}}}{{\pi h_g \kappa_{\mathrm{LA+TA}}}(1- \exp{\frac{-a^2}{r_0^2}})}\left[\frac{1}{4}-\frac{1}{4}\mathrm{Ei}\left( \frac{-a^2}{r_0^2}\right)  +\frac{r_0^2}{4 a^2}( e^{-a^2/r_0^2}-1)   \right] - \frac{D_2}{D_1} \Delta U_{1z} + \frac{Q_{\mathrm{laser}}}{D_1} 
\end{equation}
For the flexural phonon bath, the boundary resistance is assumed to be limiting the flow of heat. This implies that $U_{1z}$ is uniform over the surface of the drum and a balance of the heat fluxes at the boundary (Eq. \ref{eq:heatfluxboundary}) can be used to calculate its value. 
\begin{equation}
\begin{aligned}
\Sigma_{q,r} Q_{1z \rightarrow qr} + \Sigma_{i,j} Q_{ij \rightarrow 1z}  = \\
 (\bar{w}_{1l\rightarrow1z}  +\bar{w}_{1t\rightarrow1z} \frac{c_{1l}}{c_{1t}} ) c_{1l} \Delta U_{1l} (r=a) - ((\bar{w}_{1z\rightarrow1l} +\bar{w}_{1z\rightarrow1t}) c_{1z} + \Sigma_r \bar{w}_{1z\rightarrow2r} c_{1z}) \Delta U_{1z} = 0,
\end{aligned}
\end{equation}
\begin{equation}
\Delta U_{1z} = \frac{(\bar{w}_{1l\rightarrow1z} +\bar{w}_{1t\rightarrow1z} \frac{c_{1l}}{c_{1t}}) c_{1l} Q_{\mathrm{laser}}/D_1}{(\bar{w}_{1l\rightarrow 1z} + \bar{w}_{1t\rightarrow1z}\frac{c_{1l}}{c_{1t}})c_{1l}D_2/D_1 +  (\bar{w}_{1z\rightarrow1l} +\bar{w}_{1z\rightarrow1t}) c_{1z} + \Sigma_r \bar{w}_{1z\rightarrow2r} c_{1z}}
\end{equation}

\end{widetext}

\end{document}